\documentclass[12pt]{article}
\usepackage{amssymb,amsmath,bm,here,amssymb,epsfig,here}

\usepackage{amssymb,amsmath,amsfonts,amsbsy,graphicx}
\usepackage{color}
\usepackage{psfrag}
\usepackage{cite}

\renewcommand{\thefootnote}{\fnsymbol{footnote}}

\newcommand{\re}{{\rm Re\,}}

\newcommand{\eq}[1]{
\begin{equation}
#1 
\end{equation}
}

\newcommand{\eqn}[1]{
\begin{eqnarray}
#1 
\end{eqnarray}
}

\newcommand\ba{\begin{eqnarray}}
\newcommand\ea{\end{eqnarray}}

\newif\iffigure
\figuretrue

\begin{document}
\title{}

\title{
\begin{flushright}
\begin{minipage}{0.2\linewidth}
\normalsize
CTPU-PTC-18-01\\
EPHOU-18-001 \\
WU-HEP-18-2 \\*[50pt]
\end{minipage}
\end{flushright}
{\Large \bf 
Modulus D-term Inflation\\*[20pt] } }

\author{Kenji~Kadota$^{1,}$\footnote{
E-mail address: kadota@ibs.re.kr}, \ 
Tatsuo~Kobayashi$^{2,}$\footnote{
E-mail address: kobayashi@particle.sci.hokudai.ac.jp}, \ 
Ikumi~Saga$^{2,}$\footnote{
E-mail address: i-saga@particle.sci.hokudai.ac.jp} \ and \ 
Keigo~Sumita$^{3,}$\footnote{
E-mail address: k.sumita@aoni.waseda.jp
}\\*[20pt]
$^1${\it \normalsize 
Center for Theoretical Physics of the Universe,
Institute for Basic Science (IBS),}\\
{\it \normalsize Daejeon 34051, Korea} \\
$^2${\it \normalsize 
Department of Physics, Hokkaido University, 
Sapporo 060-0810, Japan} \\
$^3${\it \normalsize 
Department of Physics, Waseda University, 
Tokyo 169-8555, Japan} \\*[50pt]}

\date{
\centerline{\small \bf Abstract}
\begin{minipage}{0.9\linewidth}
\medskip 
\medskip 
\small
We propose a new model of single-field D-term inflation in supergravity, where the inflation is driven by a single modulus field which transforms non-linearly under the U(1) gauge symmetry. One of the notable features of our modulus D-term inflation scenario is that the global U(1) remains unbroken in the vacuum and hence our model is not plagued by the cosmic string problem which can exclude most of the conventional D-term inflation models proposed so far due to the CMB observations. 
\end{minipage}}

\begin{titlepage}
\maketitle
\thispagestyle{empty}
\clearpage
\tableofcontents
\thispagestyle{empty}
\end{titlepage}

\renewcommand{\thefootnote}{\arabic{footnote}}
\setcounter{footnote}{0}

\section{Introduction}
The cosmic inflation offers one of the most compelling paradigms for the early universe\cite{Dvali:1994ms,Copeland:1994vg}, which is supported by the cosmological observables such as the Cosmic Microwave Background (CMB) \cite{Ade:2015xua,Ade:2015lrj}. 
Many of the proposed inflation models have been studied in the framework of supergravity because of the high energy scale involved in the inflation dynamics.

Among the popular supergravity inflation scenarios are called the D-term inflation models, 
where the D-term scalar potential provides the energy density responsible 
for the inflation dynamics. 
Models of this sort boast of an evident advantage that the inflaton field does not obtain the mass of the order Hubble scale from the higher dimensional corrections (hence spoiling the slow-roll conditions, so called $\eta-$problem) from which the F-term dominated inflation models suffer.

The D-term inflation models with a U(1) gauge symmetry were originally proposed in Ref.~\cite{Halyo:1996pp, Binetruy:1996xj, Casas:1988pa}, and most of the D-term inflation models proposed so far involve the multiple field dynamics conventionally categorized as a hybrid inflation model.
The D-term hybrid models, however, encounter a serious problem in the light of the recent CMB observations due to the inevitable cosmic string formation at the end of inflation through 
the U(1) symmetry breaking \cite{Lyth:1997pf,Jeannerot:1997is,Gott:1984ef,Kaiser:1984iv,Rocher:2004my,Rocher:2004uv}. 
Despite many attempts to suppress the cosmic string contributions to the CMB observables \cite{Rocher:2006nh,Kadota:2005mt,Lin:2006xta,Seto:2005qg,Endo:2003fr,Buchmuller:2012ex,Domcke:2014zqa}, 
few viable models of the D-term hybrid inflation exist in the literature \cite{Kadota:2017dbz,Evans:2017bjs,Domcke:2017xvu,Domcke:2017rzu}.\footnote{See also for D-term hybrid inflation 
with moduli \cite{Kobayashi:2003rx}.}

The chaotic inflation models based on the D-term potential also have been discussed \cite{Kadota:2007nc,Kadota:2008pm,Nakayama:2016eqv}, where a matter charged under the U(1) gauge symmetry is identified as the inflaton and its dynamics is governed by the D-term potential. The U(1) symmetry  is broken during the inflation and is never restored afterwards, hence such a D-term chaotic inflation is free from the cosmic string problem which the conventional D-term hybrid inflation models typically suffer from. It would be worth exploring the realization of the D-term inflation scenarios even further in view of the growing cosmological data in prospect and its possible UV completion for the model consistency, and we propose, in this paper,  a new type of D-term single-field inflation where a modulus associated with the U(1) gauge symmetry plays the role of the inflaton. In clear contrast to the previously proposed D-term inflation models, the global U(1) remains unbroken in the vacuum in our scenario and our modulus D-term inflation hence is free from the cosmic string problem.
Following the model setup of our scenario, we present the detailed dynamics and consequences of our modulus D-term inflation models to illustrate the consistency with the latest CMB data. The discussion section includes a brief comment on embedding our model into superstring theories.

\section{Model}

Let us begin our discussions by presenting the setup of our model. 
In the following discussions, the reduced Planck scale is set to unity. 
We consider the supergravity model with a $U(1)$ gauge symmetry.
We denote the $U(1)$ vector multiplet by ${\cal V}$, 
whose gauge kinetic function is obtained by 
\begin{equation}
f = f_0 + \rho T.
\end{equation}
Here, $T$ denotes a modulus superfield, and 
$f_0$ as well as $\rho$ is constant, which we set real and positive for simplicity. 
The constant $f_0$ would correspond to the vacuum expectation value (VEV) of 
another modulus, which is stabilized above the inflation scale.
Here, we use the convention that we denote a chiral superfield and its lowest scalar component by the 
same letter.
The K\"ahler potential of $T$ in our model is given by 
\begin{equation}
K = -n \log (T + \bar T) 
\end{equation}
with a real constant $n$. 
Furthermore,  $T$ transforms non-linearly as 
$T \rightarrow T -m\Lambda$ with a real constant $m$ under the $U(1)$ transformation 
${\cal V} \rightarrow {\cal V} + \Lambda + \bar \Lambda$ ($\Lambda$ is a chiral superfield).
Thus, the $U(1)$ invariant K\"ahler potential is written by 
\begin{equation}
K = -n \log (T + \bar T + m{\cal V}).
\end{equation}
That induces the $T$-dependent D-term, $D= -mn/(2\tau) + \cdots$, 
where 
\eq{
\tau\equiv \re T, 
}
for the scalar component of the superfield $T$. 
Also, we assume an additional constant Fayet-Iliopoulos (FI) term, $\xi$.
That is, the D-term is written by 
\begin{equation}
D = \xi - \frac{mn}{2\tau}. 
\end{equation}
The FI term, $\xi$, would  be generated at an energy scale higher than 
the inflation scale by strong dynamics or 
VEV of another modulus, which is stabilized by a mass heavier than 
the inflation scale.
Moreover, matter fields $\phi$ charged under the $U(1)$ gauge symmetry are set 
to vanish during the inflation.

In our model $\tau$ is identified with the inflaton. 
We set the superpotential such that the D-term scalar potential is dominant 
and we can neglect the F-term scalar potential.
Then, the  inflationary dynamics is governed by the D-term potential, 
\eq{
V=\frac18\frac1{f_0+\rho \tau}\left(-2\xi+\frac{mn}\tau \right)^2. 
}
In the conventional D-term inflation, the inflation terminates in the waterfall manner 
and the $U(1)$ gauge symmetry is broken. 
Then,  cosmic strings can be produced after the inflation, which severely conflicts with the CMB observations. 
In our scenario proposed in this paper, the $U(1)$ gauge field becomes massive eating 
the axionic component of $T$.\footnote{If $\xi$ is originated by another modulus, 
its axionic part may be included in the eaten mode.}
The minima of the potential correspond to $\tau = mn/(2 \xi)$ and $\tau \rightarrow \infty$ for 
$mn/\xi > 0$, while the minimum corresponds to $\tau \rightarrow \infty$ for $mn/\xi < 0$.
At the  minimum, $\tau = mn/(2 \xi)$, the D-term vanishes.
Thus, at this vacuum, any chiral scalar fields $\phi$ do not develop VEVs and 
the global $U(1)$ symmetry remains.
The other case, $\tau \rightarrow \infty$, is a runaway vacuum, 
where the $D$-term remains finite.
Some chiral fields $\phi$ may develop their VEVs, and global symmetries would be broken.
Then, cosmic strings may be generated, but its tension may be reduced along $\tau \rightarrow \infty$.
These are important features of our scenario different from 
the others D-term inflation scenarios, 
in particular the conventional hybrid D-term inflation scenario.

We now discuss the inflation dynamics in our scenario, and let us introduce the following re-parameterizations for convenience
\eq{
A\equiv\frac{mn}{\xi},\qquad 
B\equiv\frac{f_0}{\rho},\qquad 
C\equiv\frac{\xi^2}{\rho}. \label{eq:abc}
}
Note that  all of them  are the ratios of model parameters, and, as we illustrate in the followings, this set of parameters $(A,B,C)$ can characterize the inflation dynamics.
For instance, the actual values of $(m,n)$ do not affect the inflation dynamics as long as $(A,B,C)$ stay same.
From a string-theoretical point of view, 
we expect $m, n =\mathcal O(1)$ and the precise values of $(m, n)$ of the order unity 
do not affect the discussion of our scenario. 
The scalar potential is now given by 
\eq{
V=\frac{C}{8}\frac{(-2\tau+A)^2}{\tau^2(\tau+B)},} 
and its first derivative $V_\tau =\frac{\partial V}{\partial \tau}$ consequently reads
\eq{
V_\tau=\frac{C(-2\tau+A)}{8\tau^3(\tau+B)^2}[-2A(\tau+B)-\tau(-2\tau+A)]. 
}
We can see from this expression that the scalar potential has two vacua  for $A > 0$ as mentioned above, 
i.e., $\tau = A/2$ and $\tau \rightarrow \infty$.

The slow roll parameters are calculated as 
\eqn{
\epsilon&\equiv&\frac12\left(\frac{V'}{V}\right)^2=\frac 1n\left(\frac{2A}{-2\tau+A}+\frac{\tau}{\tau+B}\right)^2,\\
\eta&\equiv&\frac{V''}{V}=\frac 2n\left[
-\frac{4A(\tau-A)}{(-2\tau+A)^2}+\frac{4A\tau}{(-2\tau+A)(\tau+B)}+\frac{\tau(\tau-B)}{(\tau+B)^2}
\right], 
}
where $V'$ and $V''$ represent the first and second derivatives of $V$ with respect to the canonically normalized 
inflaton $\tau_{can}$ defined by 
\eq{
\frac{d\tau_{\rm can}}{d\tau}\equiv\sqrt{2K_{T\bar T}}. 
}
One can see that the slow roll conditions are satisfied in some parameter regions, e.g., 
$A\ll\tau\ll B$, to be quantitatively demonstrated later. 
These slow roll parameters describe 
the spectral tilt and the tensor-to-scalar ratio as 
\eqn{
n_s&=&1-6\epsilon+2\eta,\\
r&=&16\epsilon. 
}
Note that the tilt and the tensor-to-scalar ratio are independent of $C$ which in turn can be set for the desirable fluctuation amplitude.

There are two parameter regions where we can find plausible inflationary trajectories: 
i) $A>0$ and $B,C>0$, ii) $A<0$ and $B,C>0$. 
Note that the signs of $B$ and $C$ are always same 
because the values of $f_0$ and $\xi^2$ should be positive. 
Since the physical value of $\tau$ must also be positive, the scalar potential $V$ 
contains a singularity for $B<0$ and we cannot realize the successful inflation. 
We hence will focus on the aforementioned two regions with a positive $B$. 
In Fig.~\ref{fg:poten}, 
we show the scalar potential as a function of the canonically normalized inflaton $\tau_{can}$. 
\iffigure
\begin{figure}[H]
\centering
\includegraphics[width=8cm]{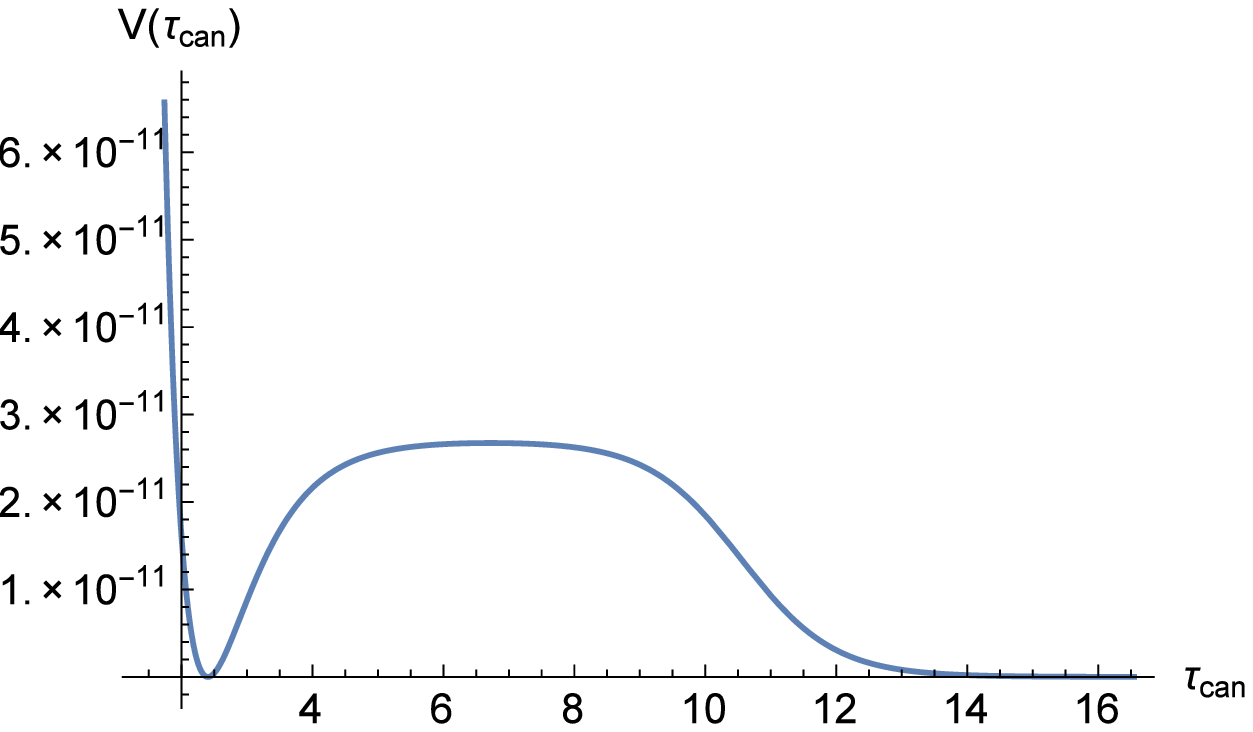}
\includegraphics[width=8cm]{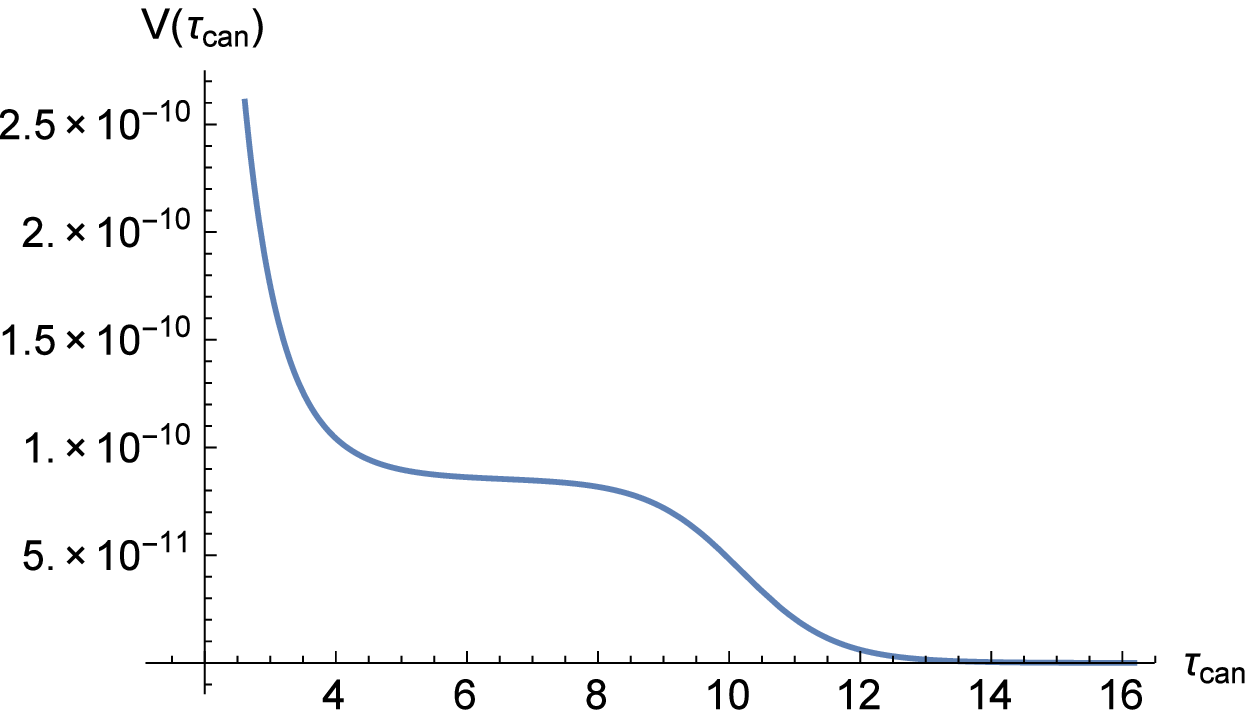}
\caption{The scalar potentials for $A>0$(left) and $A<0$(right) are shown.  }\label{fg:poten}
\end{figure} 
\fi
The parameters are set as 
$(A,\,B,\,C)=(60,\,3.0\times 10^5,\,1.6\times 10^{-4})$ 
and $(-60,\,1.8\times 10^5,\,3.1\times 10^{-4})$ 
in the left and the right panels, respectively. 
In both panels, one can find a 
flat hill suitable for the inflation. Indeed, 
the slow roll conditions, $\epsilon,\,\eta<1$, 
are satisfied around the hilltop as shown in Fig.~\ref{fg:sloro}. 
For $A > 0$, as shown in the left panel of Fig.~\ref{fg:poten}, 
there are two inflaton trajectories.
One trajectory starts from around the hilltop to the minimum $\tau = A/2$, 
and the other goes towards $\tau \rightarrow \infty$.
For $A <0$, as shown in the right panel  of Fig.~\ref{fg:poten},  
the inflaton trajectory would start from the plateau towards $\tau \rightarrow \infty$.
\iffigure
\begin{figure}[H]
\centering
\includegraphics[width=8cm]{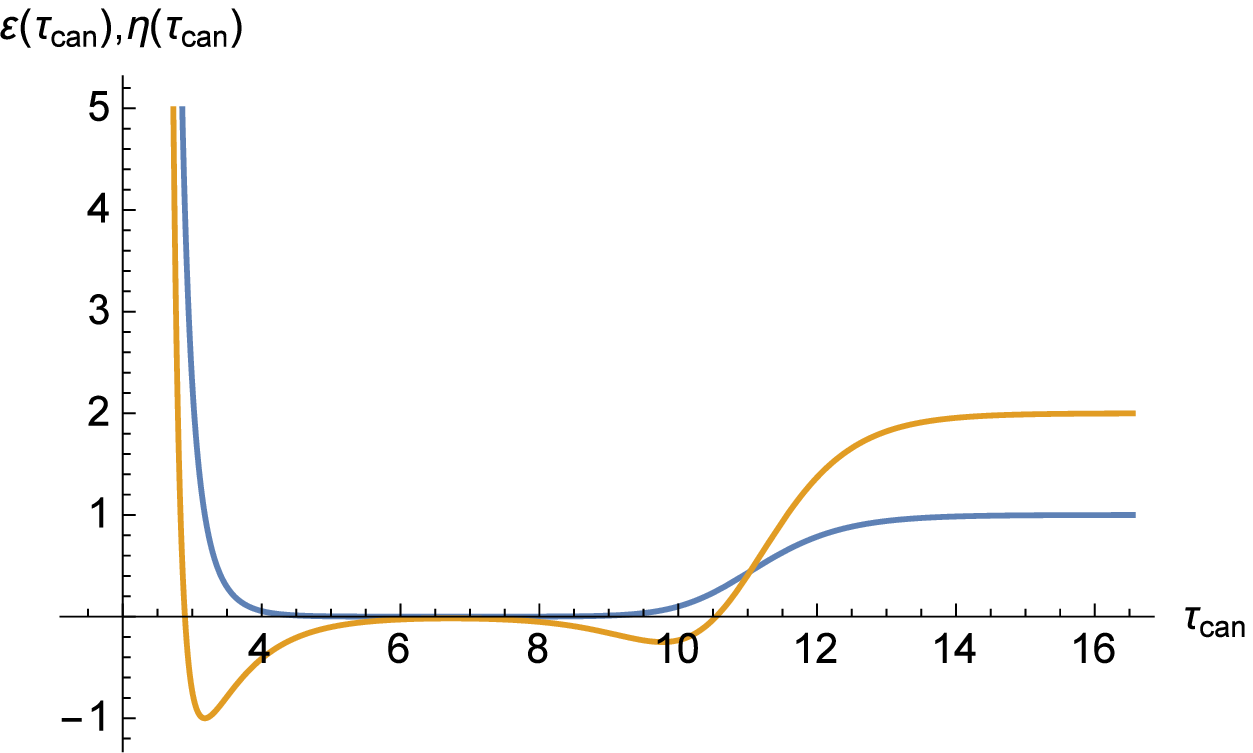}
\includegraphics[width=8cm]{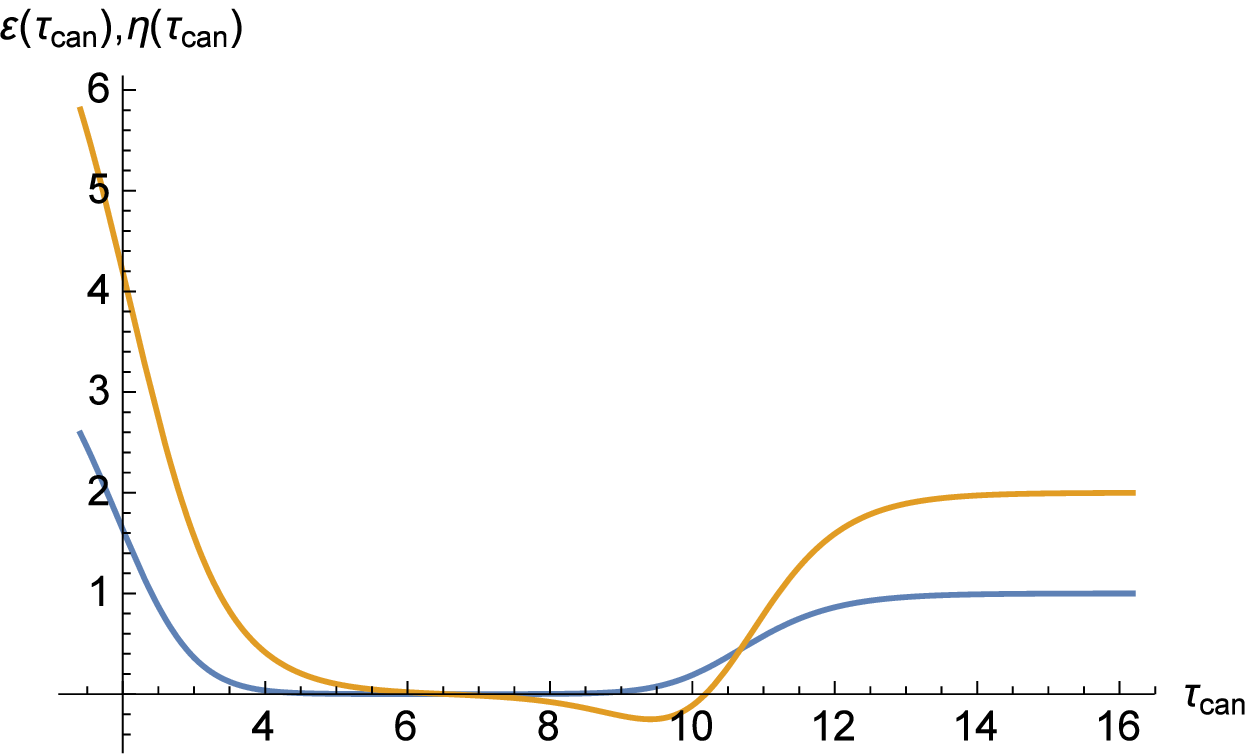}
\caption{We evaluate the slow roll parameters, $\epsilon$ (blue) and $\eta$ (orange), 
with the same values of input parameters as those used in Fig.~\ref{fg:poten}. \label{fg:sloro}
 }
\end{figure} 
\fi

\subsection{Case 1 : $A>0$}
We here study the case with $A>0$, where the typical form of the scalar potential is shown in the left panel of Fig.~\ref{fg:poten}. 
For convenience we introduce the normalized quantities
\eq{
\tilde\tau\equiv\frac{\tau}{A}, \qquad \tilde B\equiv \frac{B}{A}.\label{eq:tbtb}
}

One can see that the scalar potential becomes almost constant, 
\eq{
\frac{\xi^2}{2f_0}, 
} 
for 
\eq{
\frac12\ll\tilde\tau\ll \tilde B.
} 
In this region, the hilltop point $\tilde \tau_0$ sits around
\begin{equation}
\tilde\tau_0\sim\sqrt {\tilde B}. 
\end{equation}
The inflation terminates when the inflaton goes outside of this flat region. 
The efolding number under the slow-roll conditions reads
\eqn{
N&=&\int_{ \tilde\tau_f}^{\tilde\tau_i}\frac{V}{V_{ \tilde\tau}}\left(\frac{d\tau_{\rm can}}{d\tilde\tau}\right)^2d \tilde\tau\\
&=&-\frac{n}{8}\left[2\log \tilde\tau+\log\left|2{ \tilde\tau}^2-3 \tilde\tau-2\tilde B\right|
+\frac{8\tilde B+5}{\sqrt{9+16\tilde B}}\log\left|\frac{4 \tilde\tau-3-\sqrt{9+16\tilde B}}{4 \tilde\tau-3+\sqrt{9+16\tilde B}}
\right|\right]^{ \tilde\tau_i}_{ \tilde\tau_f}, \label{eq:efo}
}
where $[f(x)]^a_b\equiv f(a)-f(b)$ and 
$ \tilde\tau_f$ represents the end of inflation when the slow-roll condition is violated and $ \tilde\tau_i$ represents the field value $N$ efolding before the end of inflation.
For the clarity of our presentation in this section, we adopt $ \tilde\tau_f=\frac12$ when the inflaton goes 
toward the minimum, $\tilde \tau = \frac12$ ($\tau = \frac{A}{2}$), 
starting from the hilltop and $ \tilde\tau_f=\tilde B$ 
when the inflaton goes towards $\tilde \tau \rightarrow \infty$.
We checked and will show explicitly later that this simplicity does not affect the  accuracy of the estimation of $N$ by more than a few percent for the parameter range discussed in this section.

Let us first focus on the trajectory towards the minimum $ \tilde\tau_f=\frac12$. 
Since the efolding number and the spectral tilt depend only on $ \tilde\tau_i$, $\tilde B$, and $n$ 
for the fixed value of $ \tilde\tau_f=\frac12$, 
we can estimate them on $(\tilde B, \tilde\tau_i)$-plane for $n=1,2,$ and $3$. 
The results are shown in Fig.~\ref{fg:posiA}. 
\iffigure
\begin{figure}[H]
\centering
\includegraphics[width=8cm]{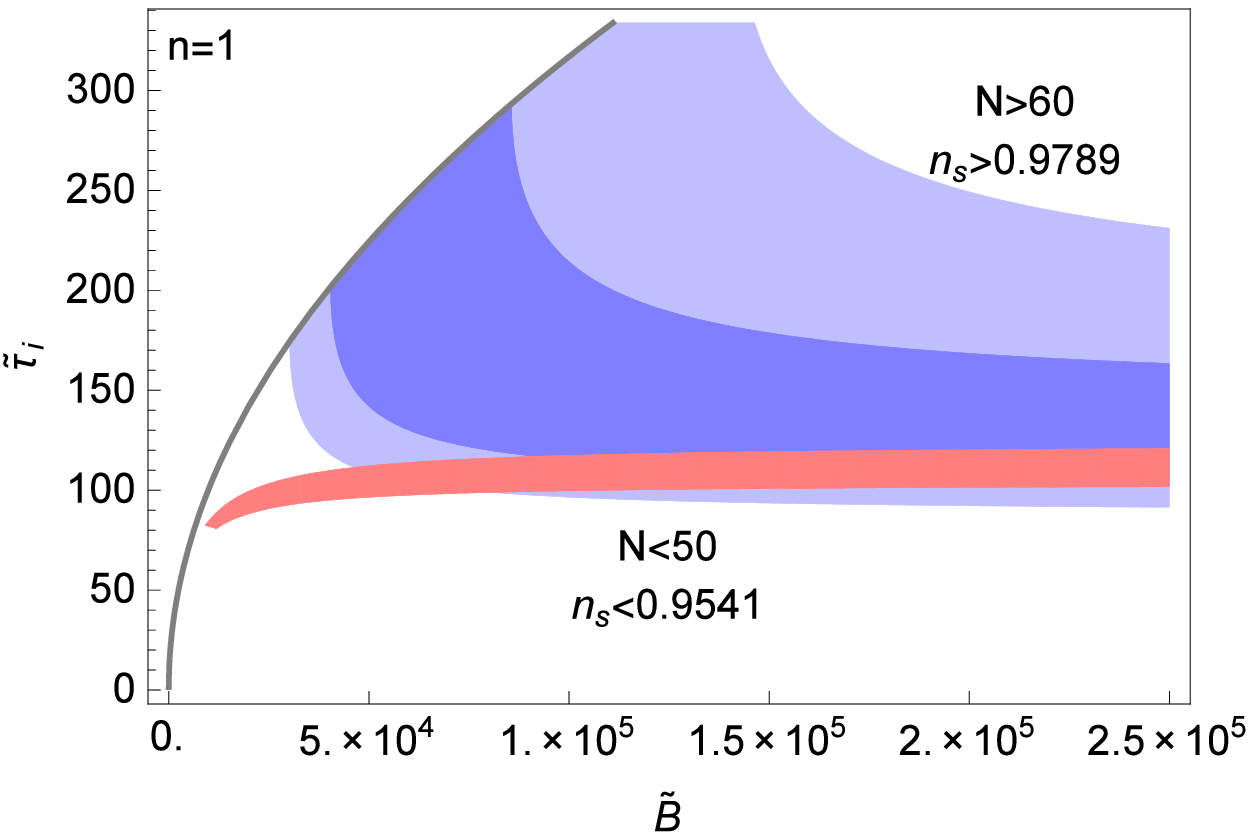}
\includegraphics[width=8cm]{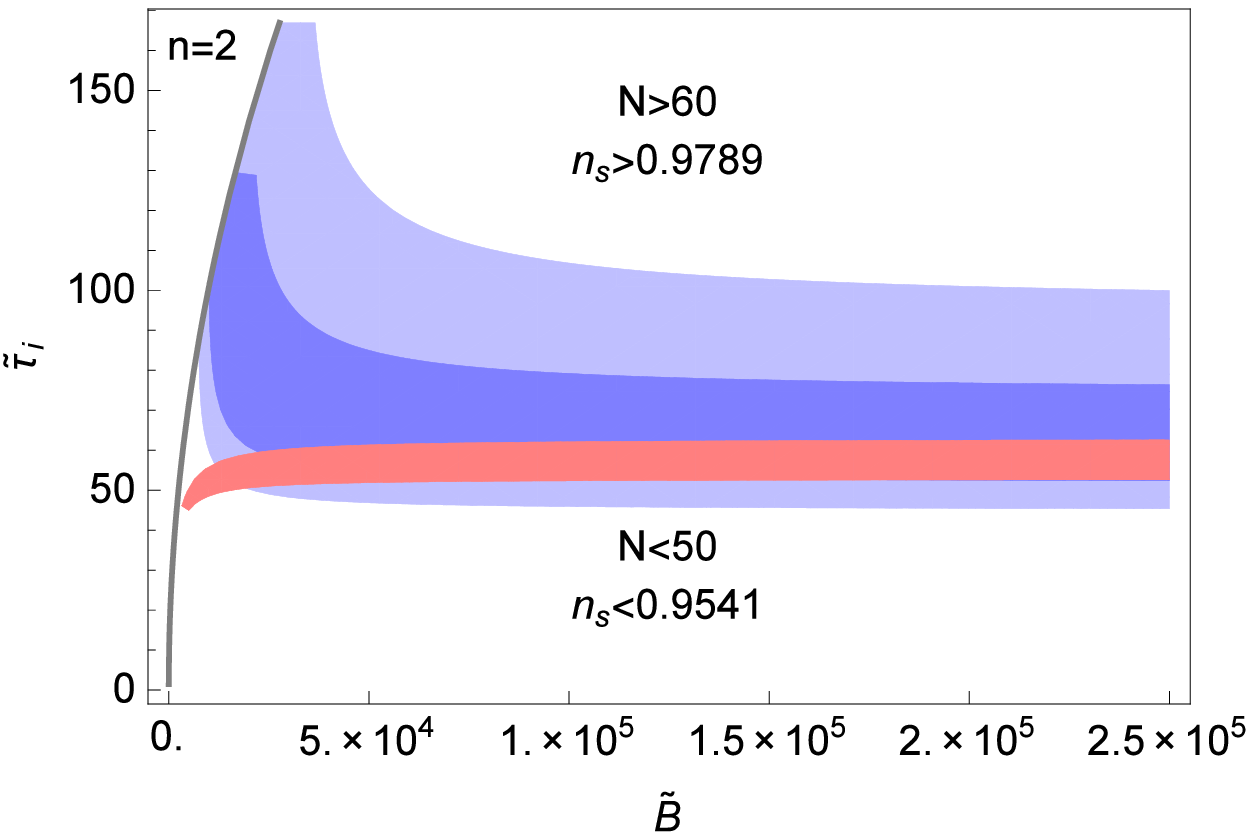}
\includegraphics[width=8cm]{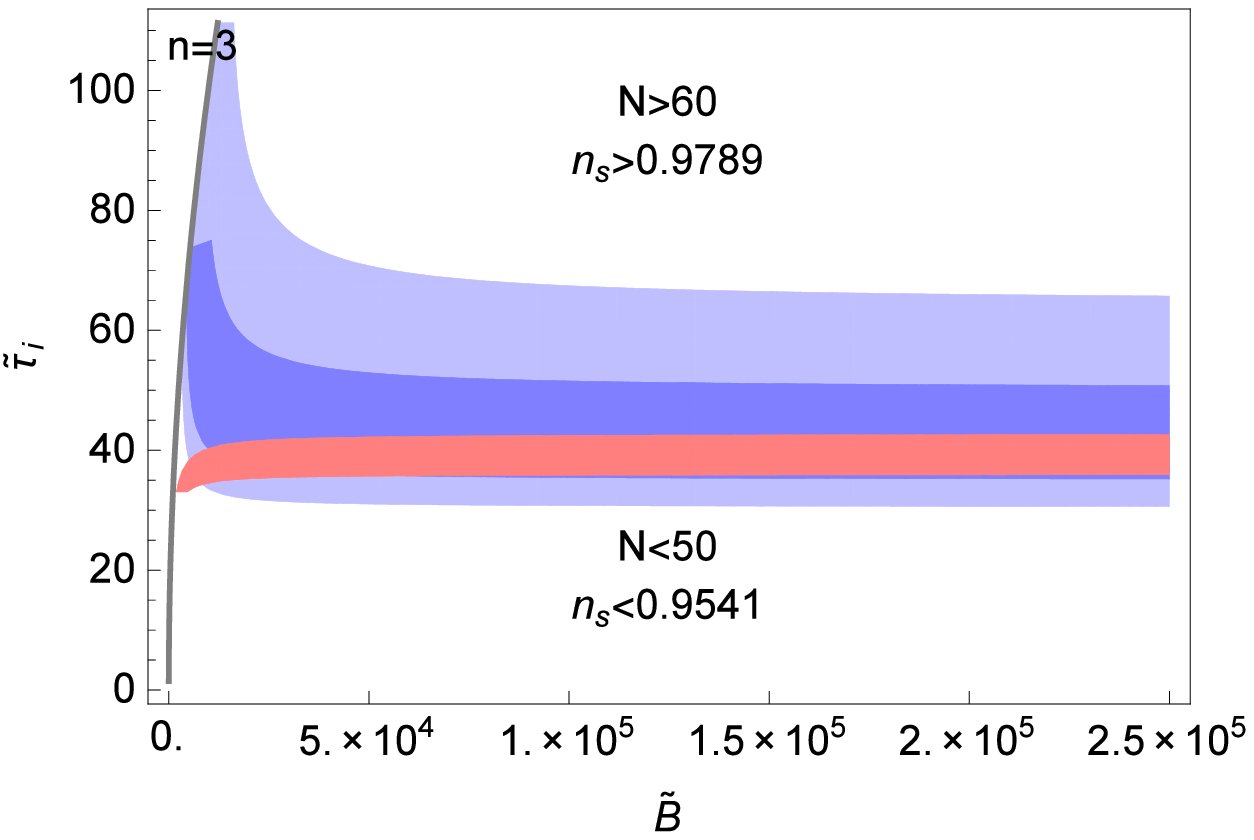}
\caption{We evaluate $n_s$ and $N$ on $(\tilde B,\,\tilde\tau_i)$-plane 
for $A>0$ when the inflaton rolls towards the minimum 
$\tilde \tau =\frac12$ ($\tau = \frac{A}{2}$). 
Three panels correspond to different values of $n$. 
The black solid line expresses $\tilde\tau_i=\tilde\tau_0$ 
($\tilde\tau_0$ represents the hilltop) and the region below this line is for the inflaton rolling 
towards the minimum $\tilde \tau =\frac12$ (the other region corresponds to the inflaton rolling 
towards $\tilde \tau \rightarrow \infty$). }\label{fg:posiA}
\end{figure} 
\fi
The deep (light) blue region corresponds to the $1\sigma$ ($2\sigma$) range of the 
CMB data on $n_s$, and we get $50\leq N\leq60$ in the red region. 

We find from these figures the following hierarchies, 
\eq{
1\ll\tilde\tau_i\ll\frac{\tilde B}{\tilde\tau_i}, \label{eq:hierar}
} 
which gives rise to some predictive results of the inflationary dynamics. 
First, this simplifies the expression for the efolding number (\ref{eq:efo}) as 
\eq{
N\approx \frac n2\tilde\tau_i. 
}
For example $\tilde\tau_i\approx100$ is predicted for $N=50$ and $n=1$, 
which is consistent with the figure. 
The expression for the tensor-to-scalar ratio is also simplified as 
\eq{
r=\frac {16}n\left(\frac{2}{-2\tilde\tau_i+1}+\frac{\tilde\tau_i}{\tilde\tau_i+\tilde B}\right)^2\approx
\frac {16}n\left(\frac1{\tilde\tau_i}\right)^{2}.\label{eq:rrr} 
}
One can estimate the typical value of $r$ as $\mathcal O(10^{-3})$. 
As we mentioned, the curvature perturbation amplitude, 
\eq{
A_s\equiv\frac{V^3}{12\pi^2(V')^2} = 
\frac{nC(2 \tilde\tau_i-1)^2}{192\pi^2 A \tilde\tau_i^2( \tilde\tau_i+\tilde B)} 
\left(\frac{ \tilde\tau_i}{ \tilde\tau_i+\tilde B}-\frac{2}{2 \tilde\tau_i-1}\right)^{-2}, 
}
can be controlled by varying the parameter $C$. 
As a result, the parameter should be determined as 
\eq{
C=
\frac{192\pi^2 A \tilde\tau_i^2( \tilde\tau_i+\tilde B)} {n(2 \tilde\tau_i-1)^2}
\left(\frac{ \tilde\tau_i}{ \tilde\tau_i+\tilde B}-\frac{2}{2 \tilde\tau_i-1}\right)^2
A_s \label{eq:das}
}
where $A_s=2.2\times 10^{-9}$. 
We have three parameters $A$, $\tilde B$ and $C$ to match three observable quantities ($A_s, n_s,r$), which can lead to a large parameter space to realize the observed CMB spectrum.

We propose the reasonable sample values of the parameters, 
focusing on a theoretical consistency with respect to the magnitude of $f_0$ and $\tau_f$. 
Starting from the parameterizations (\ref{eq:abc}), one obtains 
\eq{
f_0=\frac{m^2n^2B}{CA^2}\approx\frac{m^2n^3}{48\pi^2 A^2}\frac{\tilde\tau_i^2}{A_s}, \label{eq:apps}
}
where we used Eqs.~(\ref{eq:hierar}) and (\ref{eq:das}). 
Let us focus on the case with $m=n=1$ for simplicity. 
Fig.~\ref{fg:posiA} tells us that the $2\sigma$ data of the CMB observation 
with $N=55$ predicts $\tilde\tau_i\sim110$ and $\tilde B\gtrsim10^5$. 
The tensor-to-scalar ratio is then estimated as $r\sim 0.0013$.
Then, we can estimate $f_0\sim10^{10}/A^2$ by Eq.~(\ref{eq:apps}).
That is, we find
\begin{equation}\label{eq:app-A}
A \sim 10^5 \times f_0^{-1/2}.
\end{equation}
Recall that $A$ determines $\xi$ and the minimum $\tau_f$, 
i.e., $\xi^{-1}=2\tau_f=A$.
Here, we use $\tilde B \sim 10^{5}$ ($\tilde B = {B}/{A} = {f_0}/{(A\rho)}$) 
from Fig.~\ref{fg:posiA}, i.e., $\rho \sim 10^{-5} \times f_0/A$.
Note that the choice of $\tilde B$ affects only the magnitude of $\rho$.
Using Eq.~(\ref{eq:app-A}) we find 
\begin{equation}\label{eq:app-rho}
\rho \sim 10^{-10} \times {f_0}^{3/2}.
\end{equation}
We require that the FI-term $\xi$ should be smaller than the Planck scale, i.e.,
$\xi (=A^{-1}) \leq {\cal O}(1)$.
This leads to 
\begin{equation}
f_0 \leq 10^{10}.
\end{equation}
For example, for $f_0 \sim 10^8$, we obtain 
\begin{equation}\label{eq:f=10-8}
\rho \sim 10^2, \qquad \tau_f \sim A \sim 10,\qquad \xi \sim 10^{-1}.
\end{equation}
Similarly, we obtain 
\begin{equation}\label{eq:f=10-6}
\rho \sim 10^{-1}, \qquad \tau_f \sim A \sim 10^2,\qquad \xi \sim 10^{-2},
\end{equation}
for $f_0 \sim 10^6$,
\begin{equation}\label{eq:f=10-4}
\rho \sim 10^{-4}, \qquad \tau_f \sim A \sim 10^3,\qquad \xi \sim 10^{-3},
\end{equation}
for $f_0 \sim 10^4$,
\begin{equation}\label{eq:f=10-2}
\rho \sim 10^{-7}, \qquad \tau_f \sim A \sim 10^4,\qquad \xi \sim 10^{-4},
\end{equation}
for $f_0 \sim 10^2$.

On the other hand, when we set 
$f_0\sim\tau_f=A/2$ in Eq.~(\ref{eq:apps}), we obtain 
\eq{
A\sim\left(\frac{m^2n^3}{24\pi^2}\frac{\tilde\tau_i^2}{A_s}\right)^{1/3}.
}
For  $m=n=1$ and $\tilde B\gtrsim10^5$, 
the desirable CMB spectrum arises with 
\eq{
f_0\sim\tau_f\sim 1430,\qquad 
\xi=1/A\sim0.00035,\qquad 
\rho\lesssim 10^{-5} 
}
We can easily find the field value of $\tilde\tau$ 
at which one of the slow roll conditions is violated, $\epsilon=1$ or $\eta=1$, 
with these input values ($\rho=10^{-5}$). 
Substituting the obtained value into $\tilde\tau_f$ appearing in Eq.~(\ref{eq:efo}), 
we obtain $N=55.5$. 
This means that our estimation with $\tilde\tau_f=1/2$ for simplicity is 
reliable as mentioned before.

We will investigate the other trajectory in the rest of this subsection, where the inflaton goes 
towards $\tau \rightarrow \infty$  and 
we set $\tilde\tau_f=\tilde B$. 
Note that this runaway trajectory should be uplifted in order to 
make the modulus take its finite VEV. 
We simply assume here that the uplifting can be realized, 
for instance, by adding the positive exponential term produced 
by stringy non-perturbative effects 
to the superpotential, without affecting the inflationary dynamics\footnote{
Realization of such a positive exponent is discussed in Ref.~\cite{Abe:2005rx,Abe:2008xu}.}. Thus we will not specify the details of the uplifting mechanism here. 
Adopting $\tilde\tau_f=\tilde B$, we again evaluate the spectral tilt and the efolding number 
on $(\tilde B,\tilde\tau_i)$-plane as shown in Fig.~\ref{fg:posiAr}.

\iffigure
\begin{figure}[H]
\centering
\includegraphics[width=8cm]{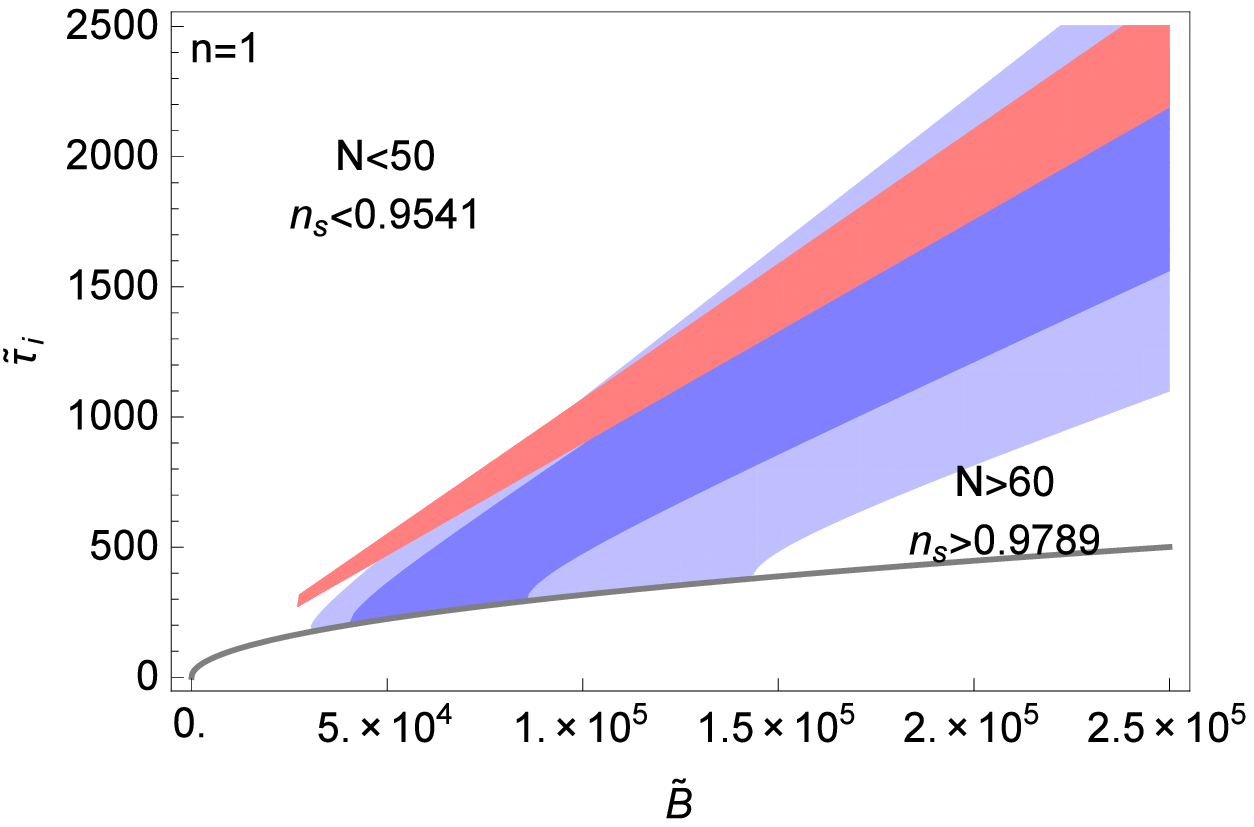}
\includegraphics[width=8cm]{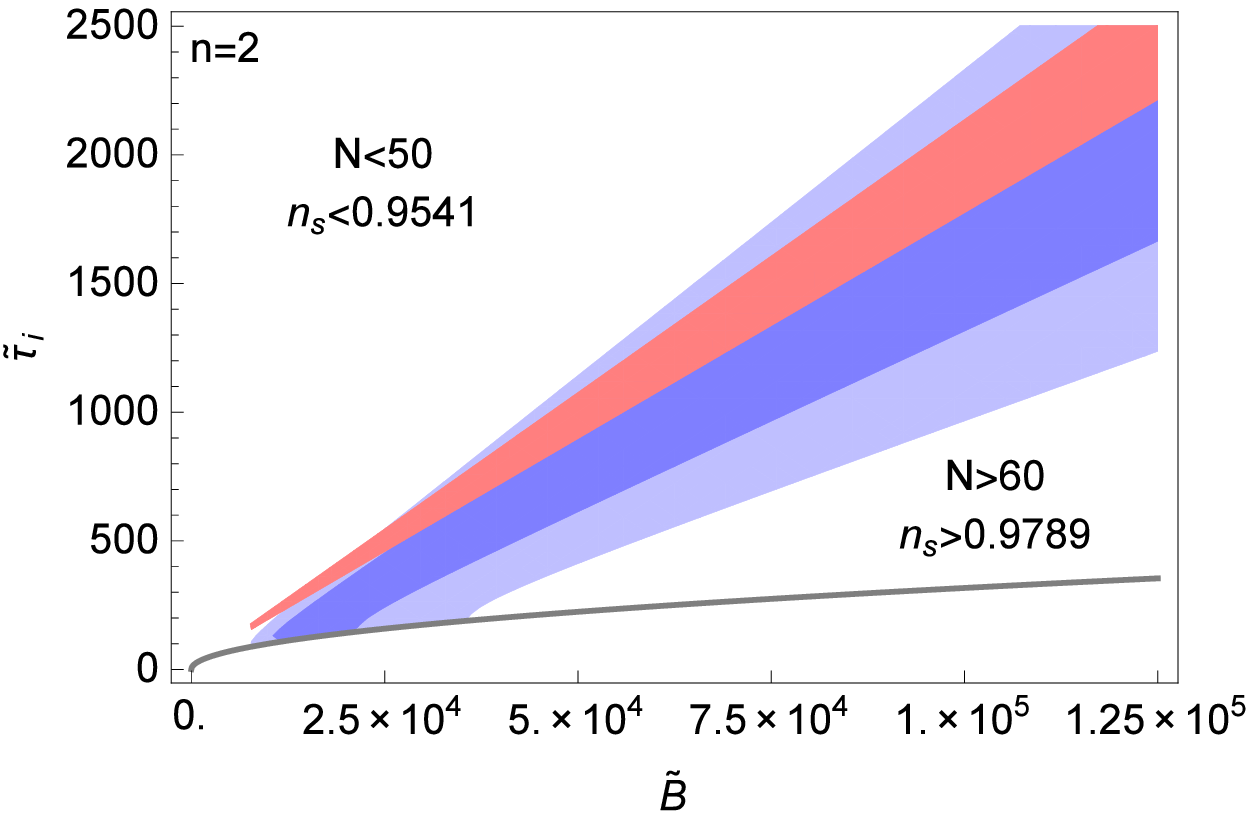}
\includegraphics[width=8cm]{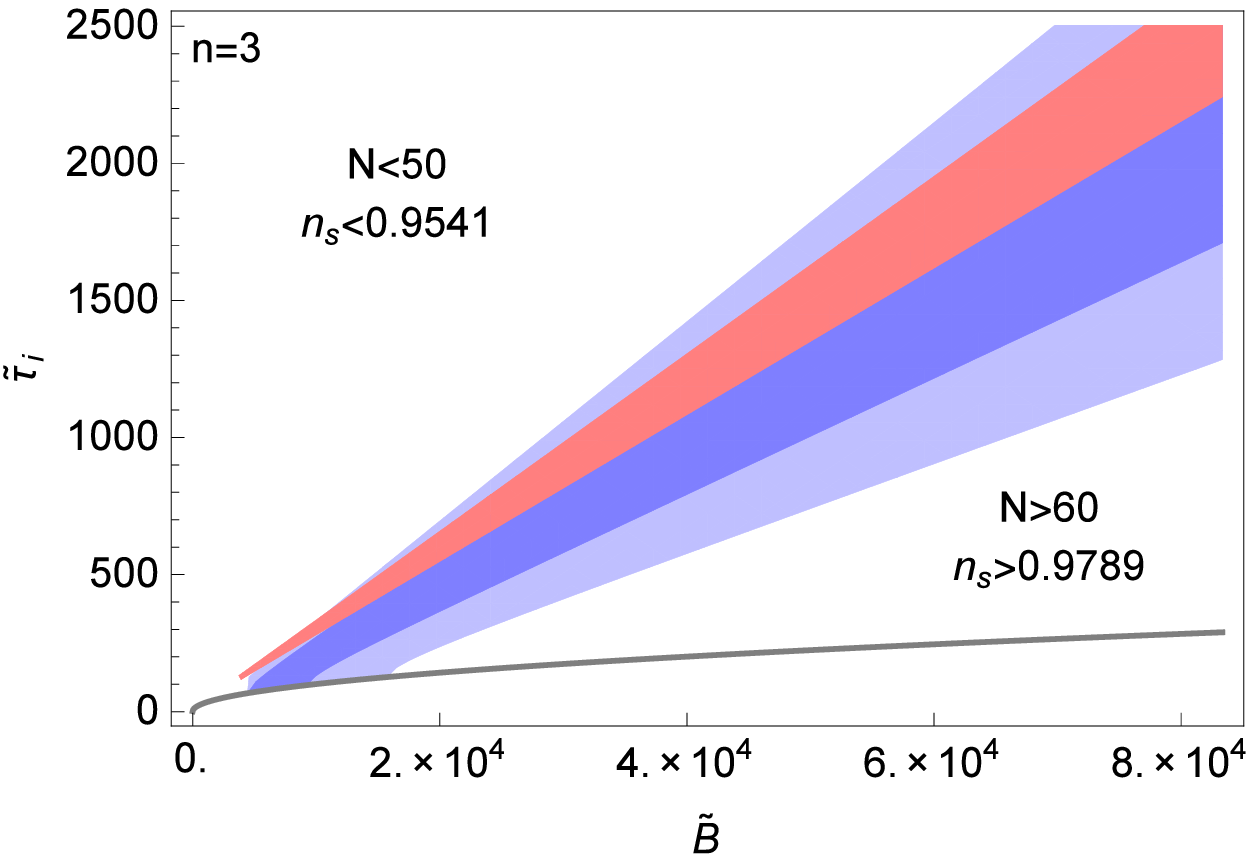}
\caption{We evaluate $n_s$ and $N$ on $(\tilde B,\,\tilde\tau_i)$-plane 
for $A>0$ when the inflaton rolls in the positive direction. 
We again adopt $n=1,2,$ and $3$ in these three panels. 
The black solid line represents $\tilde\tau_i=\tilde\tau_0$ and 
we are now focusing on a region above this line. }\label{fg:posiAr}
\end{figure} 
\fi

These figures tell us that the hierarchy 
\eq{
1\ll\frac{\tilde B}{\tilde\tau_i}\ll\tilde\tau_i \label{eq:hierar2}
}
is obtained in this trajectory. 
This is different from Eq.~(\ref{eq:hierar}) but is no surprising 
because the initial field value is now 
larger than what corresponds to the position of the hilltop $\sqrt{\tilde B}$. 
This relation modifies the expressions (\ref{eq:rrr}) and (\ref{eq:apps}) as 
\eq{
r\approx\frac{16}n\left(\frac{\tilde\tau_i}{\tilde B}\right)^2
}
and 
\eq{
f_0\approx\frac{m^2n^3}{48\pi^2 A^2}\frac{1}{A_s}\left(\frac{\tilde B}{\tilde\tau_i}\right)^2, 
}
respectively. 
The hierarchy (\ref{eq:hierar2}) leads to a tiny value of the tensor-to-scalar ratio also 
in the present case. 
Let us again focus on the case with $m=n=1$. 
When we use $\tilde \tau_i \sim 10^3$ and $\tilde B \sim 10^{5}$ from Fig.~\ref{fg:posiAr}, 
we find $f_0 \sim 10^{10}/A^2$.
This is the same relation between $A$ and $f_0$ as Eq.(\ref{eq:app-A}).
In addition, we obtain the same relation between $\rho$ and $f_0$ 
as Eq.~(\ref{eq:app-rho}) because of $\tilde B \sim 10^{5}$.
Thus, for a fixed value of $f_0$, we obtain the values of $\rho, A, \xi$ similar to 
Eqs.~(\ref{eq:f=10-8}), (\ref{eq:f=10-6}), (\ref{eq:f=10-4}), (\ref{eq:f=10-2}) except $\tau_f$.
(Note that $\tau_f$ is not determined by $A$ in the present case.)

For example, when we set $f_0=\tau_f$, we obtain 
\eq{
A=\left(\frac{m^2n^3}{48\pi^2}\frac{1}{A_s}\frac{\tilde B}{\tilde\tau_i^2}\right)^{1/3}. 
} 
Inserting this equation into the above expression for $f_0$, we obtain
\eq{
f_0=\tau_f=\left(\frac{m^2n^3}{48\pi^2}\frac{1}{A_s}\left(\frac{\tilde B}{\tilde\tau_i}\right)^2\right)^{1/3}\tilde B^{2/3}. 
}
Since we can see from Fig.~\ref{fg:posiAr} that $\tilde B/\tilde\tau_i$ is roughly 
seen as a constant in the consistent region, 
the last factor is important. 
A smaller value of $\tilde B$ is favorable in order to realize moderate values of $f_0$ and $\tau_f$. 
A lower limit on $\tilde B$ is found as $\tilde B\gtrsim10^5$. Thus we adopt $\tilde B=10^5$, 
and 
the value of $\tilde\tau_i$ is then fixed to satisfy the CMB observation 
with a proper efolding number leading to $\tilde\tau_i=1000$. 
Eventually the model parameters are determined as 
\eq{
f_0=\tau_f=4.6\times10^{6},\qquad \xi=1/A=0.0022, \qquad \rho=0.1. 
}
We again estimate the value of $\tilde\tau$ at which a slow roll condition 
is violated, in order to evaluate the efolding number accurately. 
This leads to $N = 54.7$ and verifies our expectation that our simplification 
using $N=55$ does not affect our discussions on the inflation dynamics. 

\subsection{Case 2 : $A<0$}
Next we discuss the model with a negative value of $A$. 
In this case, the viable inflationary trajectory would start from 
the plateau towards $\tau \rightarrow \infty$.
Within the range of the trajectory, 
the potential shape looks similar to the previous one with $A >0$. 
Indeed, we will obtain a similar result in the following analysis. 
First, we estimate $n_s$ and $N$ on $(\tilde B,\,\tilde\tau_i)$ 
and show that in Fig.~\ref{fg:negaAr}. 
\iffigure
\begin{figure}[H]
\centering
\includegraphics[width=8cm]{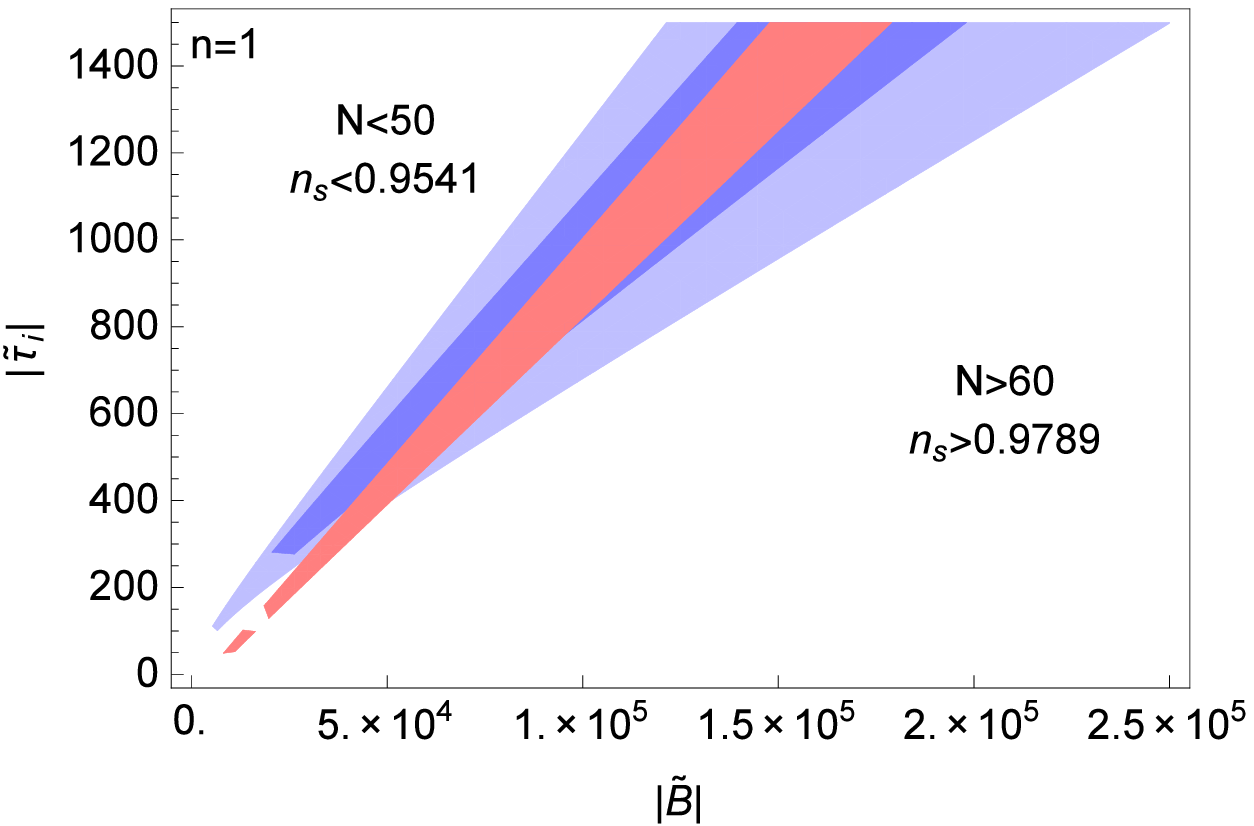}
\includegraphics[width=8cm]{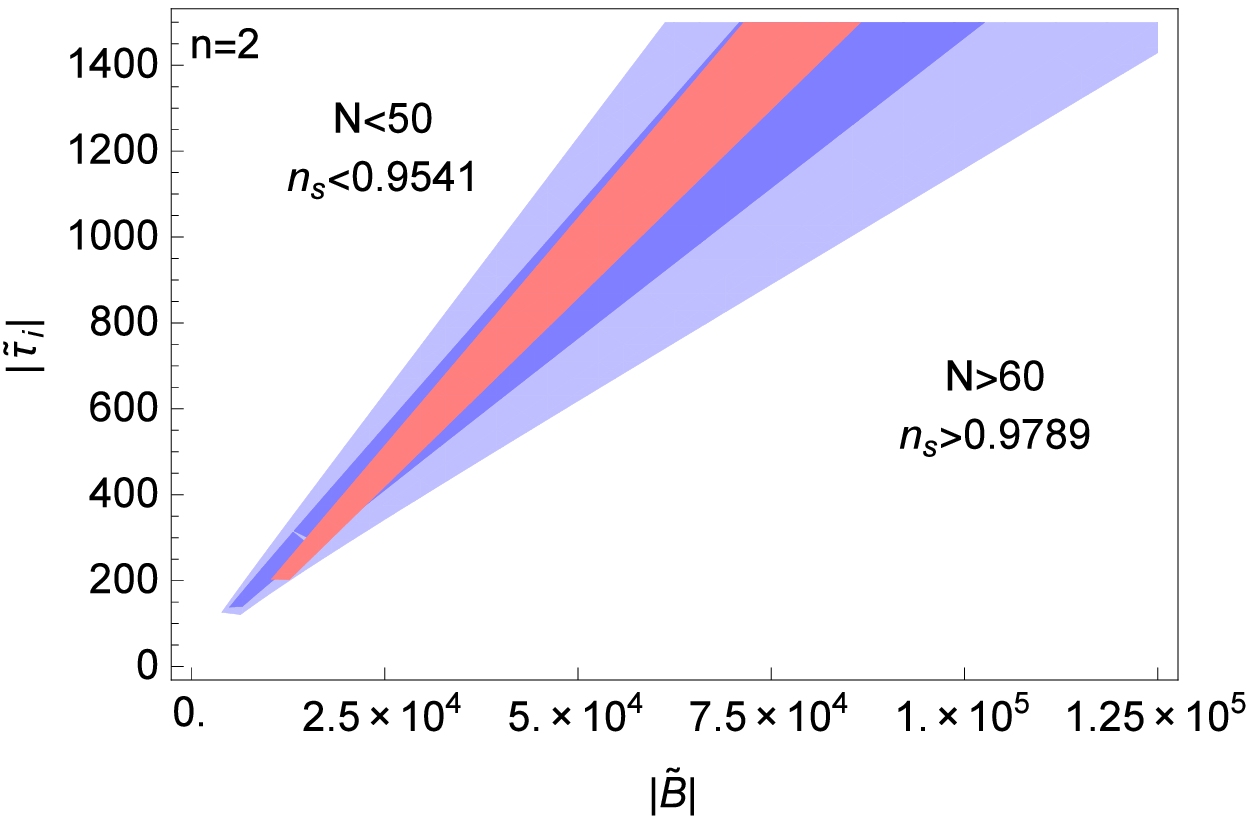}
\includegraphics[width=8cm]{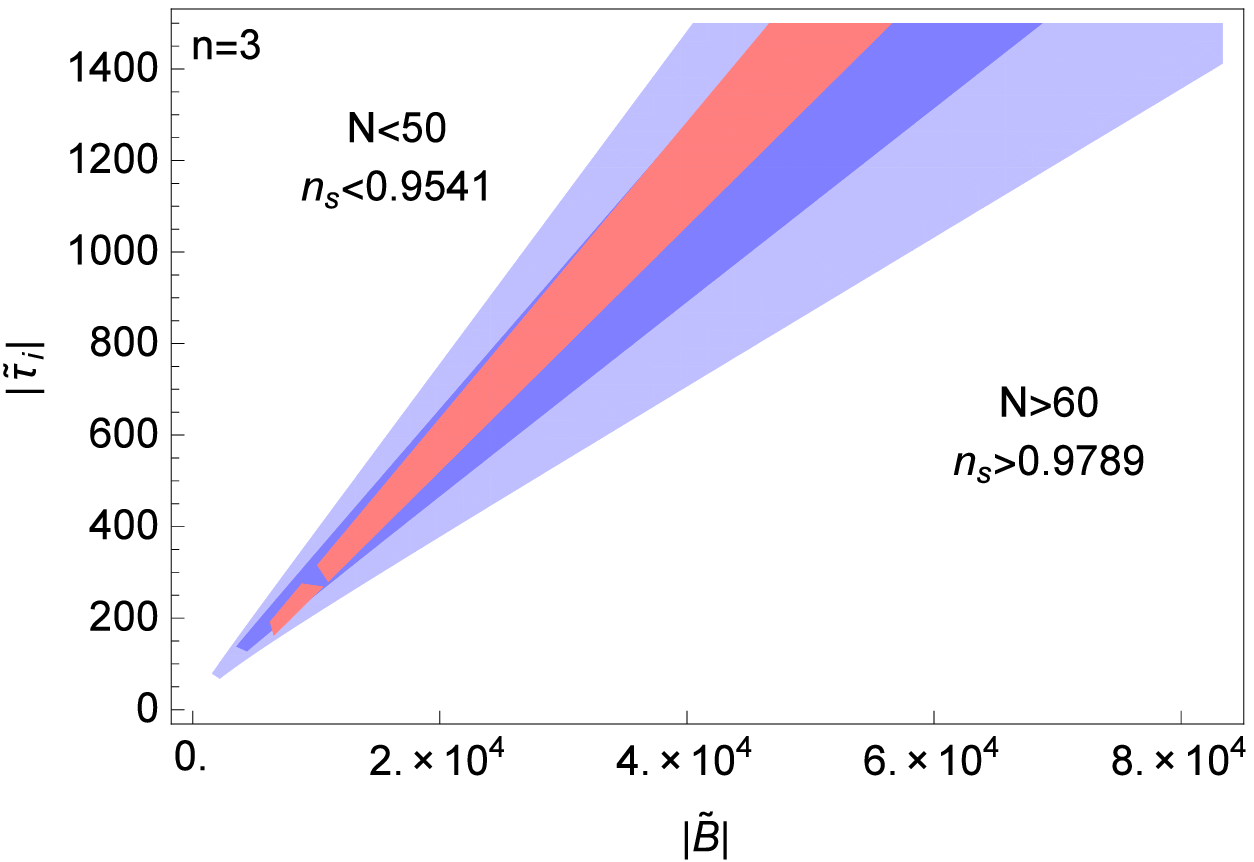}
\caption{$n_s$ and $N$ on $(\tilde B,\,\tilde\tau_i)$-plane 
for $A<0$ and $n=1,2,$ and $3$. }\label{fg:negaAr}
\end{figure} 
\fi
Although Fig.~\ref{fg:negaAr} and Fig.~\ref{fg:posiAr} look alike, 
there is a notable difference. 
Both of $\tilde B$ and $\tilde\tau_i$ are negative because of $A<0$ 
and the two axes correspond to their absolute values. 
However, the absolute values of $\tilde B$ and $\tilde\tau_i$ are similar to 
those in Fig.~\ref{fg:posiAr}.
The efolding number is now given by, whose expression differs from that for the positive $A$ given in Eq.~(\ref{eq:efo}),
\eq{
N=-\frac{n}{8}\left[2\log|\tilde\tau|+\log\left|2{ \tilde\tau}^2-3 \tilde\tau-2\tilde B\right|
+\frac{2(8\tilde B+5)}{\sqrt{-9-16\tilde B}}\tan^{-1}\left(\frac{4 \tilde\tau-3}{\sqrt{-9-16\tilde B}}\right)
\right]^{ \tilde\tau_i}_{ \tilde\tau_f}. \label{eq:efo2}
}
We find that the same hierarchy (\ref{eq:hierar2}) shows up here as well.
Therefore, the consistent CMB observables can be realized by the same parameter values as 
those for the trajectory towards $\tau \rightarrow \infty$ in the previous subsection 
except replacing $A \rightarrow -A$.

\section{Conclusion and Discussion}
We have proposed a new model of the single-field D-term inflation 
in the context of supergravity theory, introducing a $U(1)$ gauge symmetry under which a modulus field transforms nonlinearly. 
Assuming the charged matter fields decoupled or stabilized at the origin, 
the D-term potential has a plateau flat enough to realize the slow-rolling of the modulus field. 
Following the quantitative analysis, we have shown that our modulus D-term inflation leads to the inflationary trajectories consistent with the latest CMB data. 
Our modulus D-term inflation scenarios would open up a new avenue for the inflation model building in supergravity 
theory.

It is of great interest to embed our supergravity model into superstring theory, i.e., 
string-derived supergravity theory.
The modulus value would correspond to a size of the compact space.
Only if $\tau \geq {\cal O}(1)$,  the supergravity description would be reliable 
as the four-dimensional low-energy effective field theory of superstring theory.
Our parameter space shown in the previous section corresponds to 
$\tau \geq {\cal O}(1)$.
Indeed, a rather large value such as $\tau = {\cal O}(10)-{\cal O}(10^4)$ would be 
favorable.
Furthermore, the constants, $f_0$ and $\xi$, play important roles in our model.
The constant, $f_0$, would correspond to a VEV of another modulus $T'$ in string-derived supergravity theory.
In particular, $f_0$ would be linear in $T'$, 
whose VEV may be large, because $f_0$ is quite large.
The FI-term $\xi$ may be given by a VEV of modulus $T''$, e.g. $\xi = c/\re T''$, 
although $T'$ and $T''$ might be the same.
In this case, $T''$ is also quite large, because $\xi$ is very suppressed. 
Alternatively, the FI-term $\xi$ could be generated by some strong dynamics.
It would be challenging to stabilize $T'$ as well as $T^{''}$ with their masses heavier 
than the inflation scale in order to realize the desired values of $f_0$ as well as $\xi$. We will study these issues in our future work.

\section*{Acknowledgments}
We thank M. Hindmarsh for the discussions. 
K.~K. was supported by IBS under the project code IBS-R018-D1.
T.~K. was supported in part by MEXT KAKENHI Grant Number JP17H05395 and 
JSP KAKENHI Grant Number JP26247042.
K. S. was supported
by Waseda University Grant for Special Research Projects No. 2017B-253.

\end{document}